\documentclass[twocolumn,showpacs,aps,prl,superscriptaddress]{revtex4}
\usepackage{graphicx,amsmath}
\usepackage{dcolumn}

\def\figurebox#1#2#3{%
    \def\arg{#3}%
    \ifx\arg\empty
    {\hfill\vbox{\hsize#2\hrule\hbox to #2{\vrule\hfill\vbox to #1{\hsize#2\vfill}\vrule}\hrule}\hfill}%
    \else
    {\hfill\epsfbox{#3}\hfill}%
    \fi}

\RequirePackage{xspace}
\def\fb   {\ensuremath{\mbox{\,fb}}\xspace}
\def\invpb {\ensuremath{\mbox{\,pb}^{-1}}\xspace}
\newcommand{\tev}{\ensuremath{\mathrm{\,Te\kern -0.1em V}}\xspace}
\newcommand{\gev}{\ensuremath{\mathrm{\,Ge\kern -0.1em V}}\xspace}
\newcommand{\gevc}{\ensuremath{{\mathrm{\,Ge\kern -0.1em V\!/}c}}\xspace}
\newcommand{\gevcc}{\ensuremath{{\mathrm{\,Ge\kern -0.1em V\!/}c^2}}\xspace}
\def\b     {\ensuremath{b}\xspace}
\def\epem       {\ensuremath{e^+e^-}\xspace}
\def\piz   {\ensuremath{\pi^0}\xspace}
\def\taup       {\ensuremath{\tau^+}\xspace}
\def\taum       {\ensuremath{\tau^-}\xspace}
\def\tautau     {\ensuremath{\tau^+\tau^-}\xspace}
\def\Z      {\ensuremath{Z^0}\xspace}
\def\tbar  {\ensuremath{\overline t}\xspace}
\def\mumu       {\ensuremath{\mu^+\mu^-}\xspace}
\newcommand{\met}{\mbox{${E\!\!\!\!/_T}$}}
\def\pt         {\mbox{$p_T$}\xspace}
\def\Et         {\mbox{$E_T$}\xspace}
\newcommand{ \VLQ}  {\mbox{$V\!LQ3$}}
\newcommand{ \mVLQ}  {\mbox{$m_{V\!LQ3}$}}

\begin{document}

\title{Search for Third Generation Vector Leptoquarks in $p\overline{p}$
Collisions at $\sqrt{s} = 1.96 \tev$}

% CDF Jan 2007 Author List
% --------- Start of author list and affiliations ----------
\affiliation{Institute of Physics, Academia Sinica, Taipei, Taiwan 11529, Republic of China} 
\affiliation{Argonne National Laboratory, Argonne, Illinois 60439} 
\affiliation{Institut de Fisica d'Altes Energies, Universitat Autonoma de Barcelona, E-08193, Bellaterra (Barcelona), Spain} 
\affiliation{Baylor University, Waco, Texas  76798} 
\affiliation{Istituto Nazionale di Fisica Nucleare, University of Bologna, I-40127 Bologna, Italy} 
\affiliation{Brandeis University, Waltham, Massachusetts 02254} 
\affiliation{University of California, Davis, Davis, California  95616} 
\affiliation{University of California, Los Angeles, Los Angeles, California  90024} 
\affiliation{University of California, San Diego, La Jolla, California  92093} 
\affiliation{University of California, Santa Barbara, Santa Barbara, California 93106} 
\affiliation{Instituto de Fisica de Cantabria, CSIC-University of Cantabria, 39005 Santander, Spain} 
\affiliation{Carnegie Mellon University, Pittsburgh, PA  15213} 
\affiliation{Enrico Fermi Institute, University of Chicago, Chicago, Illinois 60637} 
\affiliation{Comenius University, 842 48 Bratislava, Slovakia; Institute of Experimental Physics, 040 01 Kosice, Slovakia} 
\affiliation{Joint Institute for Nuclear Research, RU-141980 Dubna, Russia} 
\affiliation{Duke University, Durham, North Carolina  27708} 
\affiliation{Fermi National Accelerator Laboratory, Batavia, Illinois 60510} 
\affiliation{University of Florida, Gainesville, Florida  32611} 
\affiliation{Laboratori Nazionali di Frascati, Istituto Nazionale di Fisica Nucleare, I-00044 Frascati, Italy} 
\affiliation{University of Geneva, CH-1211 Geneva 4, Switzerland} 
\affiliation{Glasgow University, Glasgow G12 8QQ, United Kingdom} 
\affiliation{Harvard University, Cambridge, Massachusetts 02138} 
\affiliation{Division of High Energy Physics, Department of Physics, University of Helsinki and Helsinki Institute of Physics, FIN-00014, Helsinki, Finland} 
\affiliation{University of Illinois, Urbana, Illinois 61801} 
\affiliation{The Johns Hopkins University, Baltimore, Maryland 21218} 
\affiliation{Institut f\"{u}r Experimentelle Kernphysik, Universit\"{a}t Karlsruhe, 76128 Karlsruhe, Germany} 
\affiliation{High Energy Accelerator Research Organization (KEK), Tsukuba, Ibaraki 305, Japan} 
\affiliation{Center for High Energy Physics: Kyungpook National University, Taegu 702-701, Korea; Seoul National University, Seoul 151-742, Korea; SungKyunKwan University, Suwon 440-746, Korea} 
\affiliation{Ernest Orlando Lawrence Berkeley National Laboratory, Berkeley, California 94720} 
\affiliation{University of Liverpool, Liverpool L69 7ZE, United Kingdom} 
\affiliation{University College London, London WC1E 6BT, United Kingdom} 
\affiliation{Centro de Investigaciones Energeticas Medioambientales y Tecnologicas, E-28040 Madrid, Spain} 
\affiliation{Massachusetts Institute of Technology, Cambridge, Massachusetts  02139} 
\affiliation{Institute of Particle Physics: McGill University, Montr\'{e}al, Canada H3A~2T8; and University of Toronto, Toronto, Canada M5S~1A7} 
\affiliation{University of Michigan, Ann Arbor, Michigan 48109} 
\affiliation{Michigan State University, East Lansing, Michigan  48824} 
\affiliation{University of New Mexico, Albuquerque, New Mexico 87131} 
\affiliation{Northwestern University, Evanston, Illinois  60208} 
\affiliation{The Ohio State University, Columbus, Ohio  43210} 
\affiliation{Okayama University, Okayama 700-8530, Japan} 
\affiliation{Osaka City University, Osaka 588, Japan} 
\affiliation{University of Oxford, Oxford OX1 3RH, United Kingdom} 
\affiliation{University of Padova, Istituto Nazionale di Fisica Nucleare, Sezione di Padova-Trento, I-35131 Padova, Italy} 
\affiliation{LPNHE, Universite Pierre et Marie Curie/IN2P3-CNRS, UMR7585, Paris, F-75252 France} 
\affiliation{University of Pennsylvania, Philadelphia, Pennsylvania 19104} 
\affiliation{Istituto Nazionale di Fisica Nucleare Pisa, Universities of Pisa, Siena and Scuola Normale Superiore, I-56127 Pisa, Italy} 
\affiliation{University of Pittsburgh, Pittsburgh, Pennsylvania 15260} 
\affiliation{Purdue University, West Lafayette, Indiana 47907} 
\affiliation{University of Rochester, Rochester, New York 14627} 
\affiliation{The Rockefeller University, New York, New York 10021} 
\affiliation{Istituto Nazionale di Fisica Nucleare, Sezione di Roma 1, University of Rome ``La Sapienza," I-00185 Roma, Italy} 
\affiliation{Rutgers University, Piscataway, New Jersey 08855} 
\affiliation{Texas A\&M University, College Station, Texas 77843} 
\affiliation{Istituto Nazionale di Fisica Nucleare, University of Trieste/\ Udine, Italy} 
\affiliation{University of Tsukuba, Tsukuba, Ibaraki 305, Japan} 
\affiliation{Tufts University, Medford, Massachusetts 02155} 
\affiliation{Waseda University, Tokyo 169, Japan} 
\affiliation{Wayne State University, Detroit, Michigan  48201} 
\affiliation{University of Wisconsin, Madison, Wisconsin 53706} 
\affiliation{Yale University, New Haven, Connecticut 06520} 
\author{T.~Aaltonen}
\affiliation{Division of High Energy Physics, Department of Physics, University of Helsinki and Helsinki Institute of Physics, FIN-00014, Helsinki, Finland}
\author{A.~Abulencia}
\affiliation{University of Illinois, Urbana, Illinois 61801}
\author{J.~Adelman}
\affiliation{Enrico Fermi Institute, University of Chicago, Chicago, Illinois 60637}
\author{T.~Affolder}
\affiliation{University of California, Santa Barbara, Santa Barbara, California 93106}
\author{T.~Akimoto}
\affiliation{University of Tsukuba, Tsukuba, Ibaraki 305, Japan}
\author{M.G.~Albrow}
\affiliation{Fermi National Accelerator Laboratory, Batavia, Illinois 60510}
\author{S.~Amerio}
\affiliation{University of Padova, Istituto Nazionale di Fisica Nucleare, Sezione di Padova-Trento, I-35131 Padova, Italy}
\author{D.~Amidei}
\affiliation{University of Michigan, Ann Arbor, Michigan 48109}
\author{A.~Anastassov}
\affiliation{Rutgers University, Piscataway, New Jersey 08855}
\author{K.~Anikeev}
\affiliation{Fermi National Accelerator Laboratory, Batavia, Illinois 60510}
% ...
\collaboration{CDF Collaboration\footnote{With visitors from $^a$University of Athens, 15784 Athens, Greece, 
$^b$University of Bristol, Bristol BS8 1TL, United Kingdom, 
$^c$University Libre de Bruxelles, B-1050 Brussels, Belgium, 
$^d$Cornell University, Ithaca, NY  14853, 
$^e$University of Cyprus, Nicosia CY-1678, Cyprus, 
$^f$University College Dublin, Dublin 4, Ireland, 
$^g$University of Edinburgh, Edinburgh EH9 3JZ, United Kingdom, 
$^h$University of Heidelberg, D-69120 Heidelberg, Germany, 
$^i$Universidad Iberoamericana, Mexico D.F., Mexico, 
$^j$University of Manchester, Manchester M13 9PL, England, 
$^k$Nagasaki Institute of Applied Science, Nagasaki, Japan, 
$^l$University de Oviedo, E-33007 Oviedo, Spain, 
$^m$University of London, Queen Mary College, London, E1 4NS, England, 
$^n$University of California Santa Cruz, Santa Cruz, CA  95064, 
$^o$Texas Tech University, Lubbock, TX  79409, 
$^p$University of California, Irvine, Irvine, CA  92697, 
$^q$IFIC(CSIC-Universitat de Valencia), 46071 Valencia, Spain. 
}}
\noaffiliation
% --------- End of author list and affiliations ----------

\date{\today}

\begin{abstract}
We describe a search for a third generation vector leptoquark (\VLQ)
that decays to a $b$ quark and tau lepton using the CDF II detector
and $322 \invpb$ of integrated luminosity from the Fermilab Tevatron.
Vector leptoquarks have been proposed in many extensions of the standard
model (SM).  Observing a number of events in agreement with SM expectations,
assuming Yang-Mills (minimal) couplings, we obtain the most stringent upper
limit on the \VLQ\ pair production cross section of $344 \fb$ ($493 \fb$)
and lower limit on the \VLQ\ mass of $317 \gevcc$ ($251 \gevcc$) at
$95\%$ C.L.
\end{abstract}

% PACS, the Physics and Astronomy Classification Scheme.
\pacs{14.80.-j, 13.85.Rm}

\maketitle

%%%%%%%%%%%%%%%%%%%%%%%%%%%%%%%%%%%%%%%%%%%%%%%%%%%%%%%%%%%
\newpage

Despite its extraordinary success, the standard model (SM) of elementary
particles has structural deficiencies.  The parallels between the
families of quarks and leptons suggest a possible link between these two
sectors at higher mass scales.  Leptoquarks, therefore, have been proposed
as fractionally-charged color-triplet bosons carrying both lepton and
baryon quantum numbers.  Leptoquarks appear in a wide range of theories,
including SU(5) grand unification~\cite{SU5}, superstrings~\cite{Super},
SU(4) Pati-Salam~\cite{SU4}, and compositeness models~\cite{Comp}; direct
searches for leptoquarks at hadron colliders probe the parameter space of
some of these models.

The various leptoquark states are classified according to the quantum
numbers of SM gauge group interactions~\cite{Buchmuller}.  At the Tevatron
collider, these states would be predominately pair produced through quark
anti-quark annihilation.  In general, larger cross sections, and thus better
search sensitivities, are predicted for vector (spin 1) than for scalar
leptoquarks~\cite{HewettVector}.

We search for third generation vector leptoquark (\VLQ) pair production,
and assume each \VLQ\ decays promptly to a $b$ quark and a tau lepton.
As the trilinear and quartic couplings between vector leptoquarks
and gluons can have model-dependent ``anomalous'' contributions, we examine
two scenarios:  one with Yang-Mills couplings, where vector leptoquarks
are fundamental gauge bosons of an extended gauge group, and the other
with minimal anomalous couplings~\cite{HewettVector,Blumlein}.
Previous \VLQ\ searches have been carried out in experiments at
$p\overline{p}$, $\epem$, and $ep$ colliders~\cite{CDFrunI,D0runI,OPAL,H1}.
Our new results substantially extend the reach beyond the previous
limits.

The results reported in this Letter are obtained from data corresponding
to $322 \invpb$ of integrated luminosity collected between March 2002 and
August 2004 by the CDF II detector~\cite{CDFdetector}, operating at the
Tevatron $p\overline{p}$ collider.  Detector components relevant to this
search are described briefly here.  The charged particle tracking volume,
which is inside a uniform $1.4~{\rm T}$ magnetic field, includes a
multi-layer silicon microstrip detector system and an open-cell drift
chamber (COT) that provide position, momentum, and charge information in the
range $|\eta| < 1.0$~\cite{coords}.  Outside the solenoid, electromagnetic
(EM) and hadronic calorimeters are arranged in a projective tower geometry
to measure electron, photon, and jet energies.
A set of strip and wire chambers (CES) is embedded in the EM calorimeter,
at the depth where the longitudinal development of EM showers is expected
to be maximal, and provides information used
for the identification of electron candidates and reconstruction of $\piz$
candidates that decay through $\piz\to\gamma\gamma$.  Outside
the calorimeters, drift chambers and scintillators provide muon candidate
identification in the range $|\eta| < 1.0.$

This search assumes a branching ratio ${\cal B}(\VLQ \to\b\tau)=1$, and
considers a signature where the decay products of the \VLQ\ pair,
$\taup\taum\b\overline{b}$, yield two jets from the $b$ quarks, an
electron or muon from a leptonically decaying tau, and a hadronically
decaying tau ($\tau_h$).  We do not attempt to identify the jets as
originating from $b$ quarks, as we find this would degrade the
search sensitivity.  A three-level trigger system selects events with lepton
candidates and charged tracks~\cite{LeptonTrackTrigger}.  These events
are then classified as $e\tau_h$ or $\mu\tau_h$ based on the flavor of
the leptonic tau decay.

% Particle Selection

Selected events are required to contain at least one well-identified
electron (muon) candidate that passes fully through the fiducial volume of
the COT, with transverse energy (momentum)
$\Et > 10 \gev$ ($\pt > 10 \gevc$)~\cite{coords}.  To reduce the background
due to multi-jet quantum chromodynamics (QCD) events, an isolation
requirement is imposed upon the electron or
muon candidate.  Specifically, the sum of the $\pt$ of all additional
tracks within a cone in $\eta-\phi$ space of
$\Delta R < 0.4$ around the track direction of the candidate is required to
be less than
$2 \gevc$, where $\Delta R \equiv \sqrt{ \Delta\eta^2 + \Delta\phi^2 }$.

Hadronic tau candidates are formed by matching narrow clusters of
calorimeter towers with tracks.  The procedure is described briefly here
and in detail elsewhere~\cite{ztautauprd}.  A cluster is seeded by a tower
with $\Et > 6 \gev$.  The highest $\pt$ track with $\pt > 6 \gevc$ that
points to the cluster is defined as the seed track.  Other tracks are added
if they are within an angle of $0.05$ to $0.17$ radians from the seed track,
where the angle depends on the total cluster energy.  Tau candidates with
one or three tracks are considered.  We reconstruct $\piz$ candidates as
single narrow strip and wire clusters in the CES, and include them as part
of the $\tau_h$ candidate if they are within $0.17$ radians of the seed
track and have $\Et > 1 \gev$.  The $\tau_h$ candidates are required to have
$|\eta| < 1.0$ and $\Et > 15 \gev$.  To reduce instances where a jet is
misidentified as a $\tau_h$, we place requirements on the mass formed by
constituent tracks, $m_{\rm trk} < 1.8 \gevcc$, and the mass formed from
tracks and $\piz$ mesons within the tau candidate,
$m_{{\rm trk}+\piz} < 2.5 \gevcc$.  To better discriminate hadronic taus
from electrons, we require the ratio of $\Et$ deposited in the
hadronic calorimeter to the sum of the magnitudes of the transverse momenta
of the tau tracks to be greater than $0.1$.  To obtain isolated $\tau_h$
candidates, we require that the region outside the track selection cone,
but inside a cone of $0.52$ radians from the seed track, contains no tracks
with $\pt > 1 \gevc$.  In addition, the region outside the $\piz$ inclusion
cone, but inside a cone of $0.52$ radians from the seed track, is required
to contain less than $0.6 \gev$ of summed $\Et$ due to $\piz$ candidates.

To ensure efficient event reconstruction, the electron or muon candidate
($\ell$) direction must be separated from the tau candidate direction by
$\Delta R(\tau_h,\ell) > 0.7$.  Jet candidates,
with $\Et > 15 \gev$, are identified in the region $|\eta| < 2.4$ and are
required to be separated from the lepton candidates by
$\Delta R(\ell/\tau_h,{\rm jet}) > 0.8$.

% Backgrounds

There are a number of SM processes which can mimic the \VLQ\ signal.
The first category consists of background processes which contain a real
$e\tau_h$ or $\mu\tau_h$ plus two jets.  The primary processes are
$\Z/\gamma^*\to\tau\tau$ plus two jets, and $t\tbar\to WbWb$, where one
$W$ yields a hadronic tau via $W\to\tau\nu_{\tau}$, the other $W$ similarly
yields an electron (muon) or leptonically decaying tau, and the two $b$
quarks give jets.

The second category of backgrounds consists of those that include
misidentified final state particles.  These include
$t\tbar\to WbWb$, where a jet from a hadronic W decay can be
misidentified as an electron (muon).  The processes $t\tbar\to WbWb$,
$\Z/\gamma^*\to\epem$ plus jets, and $\Z/\gamma^*\to\mumu$ plus jets also
contribute as backgrounds when an electron (muon) or jet is misidentified as
the $\tau_h$.  Events with $W$ plus jets can pass the selection if one
of at least three jets is misidentified as the $\tau_h$.
Contributions from diboson channels ($WW$, $WZ$, and $ZZ$) plus jets are
negligible.  The above contributions and their uncertainties
are estimated using {\sc pythia}~\cite{pythia} Monte Carlo simulation
and {\sc geant}~\cite{geant} CDF II detector simulation.  Background from
multi-jet QCD can contribute when jets from quarks are misidentified
as an electron (muon) or $\tau_h$.  Photon plus jets background enters
when high-$\pt$ photons convert within the detector and at least one of the
resulting electrons appears as a primary electron candidate, while
a jet is misidentified as the $\tau_h$.  Contributions from both of these
sources are estimated directly from the data, using methods described
elsewhere~\cite{ztautauprd}.

% Event Selection

Further event selection reduces the backgrounds.  Backgrounds associated
with misidentification are reduced by approximately one-half through a
requirement of oppositely charged electron (muon) and $\tau_h$ candidates,
where the charge of the $\tau_h$ candidate is defined as the sum of the
charges of all constituent tracks.  Events consistent with photon conversions
and cosmic rays are removed.  To reduce contributions from $Z$ boson
production, events are rejected if $76 < m(\ell,X) < 106 \gevcc$, where
$X$ can be the tau candidate or a second electron candidate
in the $e\tau_h$ channel, or a second muon candidate in
the $\mu\tau_h$ channel.  We require $\met > 10 \gev$, where $\met$ is
the magnitude of the missing transverse energy~\cite{coords}.
This requirement reduces backgrounds from $\Z/\gamma^*\to\tautau$ and
multi-jet QCD processes and is nearly $100\%$ efficient for the signal process.
We define $H_T$ as the scalar sum of electron (muon) candidate $\Et$,
tau candidate $\Et$, event $\met$, and the transverse energies of the
two highest $\Et$ jet candidates.  The requirements on $H_T$ are given below.
The final selection requirement is that the event must contain
two or more jets.

To simulate \VLQ\ pair production and decay~\cite{lq3simulation}, we have
added the production and decay processes to the {\sc grace}~\cite{grace}
matrix element event generator, which calculates amplitudes, and to the
{\sc gr@ppa}~\cite{grappa} interface, which speeds up computations of the 
interactions of the primary hadrons.  In addition to providing the theoretical
cross section, these programs yield events that are processed through
{\sc tauola}~\cite{tauola} to simulate tau decays,
{\sc pythia}~\cite{pythia} to simulate parton showering, fragmentation,
and additional particle decays, and {\sc geant}~\cite{geant}
for the full CDF II detector simulation.  For the first time, this framework 
includes helicity amplitudes for the full matrix element at tree
level and propagation of helicity information from the
leptoquarks to the tau decay products.  We use the parton distribution
functions (PDFs) {\sc cteq5l}~\cite{cteq5L} and renormalization energy scale
$Q^2 = \mVLQ^2$.

We determine total selection efficiencies, including factors for triggering,
geometrical and kinematic acceptance, particle candidate identification
and isolation, and background suppression criteria.  The total efficiencies,
averaged between the $e\tau_h$ and $\mu\tau_h$ channels, range from
about $2.2\%$ ($1.4\%$) to about $6.1\%$ ($5.9\%$) for Yang-Mills (minimal)
couplings over the mass range $160 \gevcc$ to $400 \gevcc$.
For the example of $\mVLQ = 320 \gevcc$ and Yang-Mills
couplings, the efficiencies for the $e\tau_h$ and $\mu\tau_h$ channels
are $(6.0 \pm 0.1)\%$ and $(6.1 \pm 0.1)\%$ respectively.

\begin{figure}[!ht]
\begin{center}
\includegraphics[width=8.5cm,angle=0]{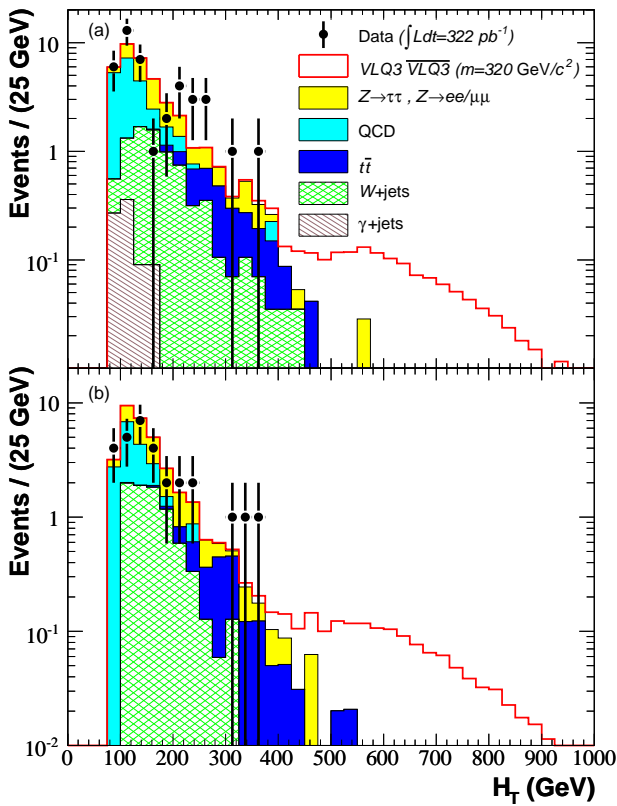}
\caption{Distributions of $H_T$ for the (a) $e\tau_h$ channel and (b)
$\mu\tau_h$ channel, including one of the control regions used for
validation ($80 < H_T < 250 \gev$), and the regions used for the search
($H_T > 250 \gev$).  The individual background contributions are shown,
as is a hypothetical signal with mass $\mVLQ = 320 \gevcc$, normalized
to the data sample size.
}
\label{fig:ht}
\end{center}
\end{figure}

We define two signal regions.  In addition, three side-band control regions
in the plane of the number of jets ($N_{\rm jets}$) versus $H_T$ are used to
verify the expected composition of the backgrounds and the distributions for
kinematic quantities.  The primary signal region (${\rm SR}_{\rm A}$)
has $N_{\rm jets} \ge 2$ and $H_T > 400 \gev$, and is sensitive to the
highest mass leptoquarks.  The secondary signal region (${\rm SR}_{\rm B}$)
has $N_{\rm jets} \ge 2$ and $250 < H_T < 400 \gev$, and adds sensitivity to
lower \VLQ\ masses (down to the previous mass limits).  The
three control regions are called
CR0J, CR1J, and CR2J, where 0J, 1J, or 2J specifies the number of jets ($0$,
$1$, or $\ge 2$).  Regions CR0J and CR1J include the $H_T$ range
$H_T > 80 \gev$, while region CR2J is restricted to $80 < H_T < 250 \gev$.
Table~\ref{table:yields} shows the expected background contributions in the
control and signal regions, as well as the number of events observed in all
regions.  The signal regions are examined only after an {\it a priori}
optimization of the $H_T$ ranges that maximizes signal sensitivity.
Control region CR1J in the $\mu\tau_h$ channel contains the largest difference
between the expected and observed number of events, with a difference of
$1.9$ sigma.  Figure~\ref{fig:ht} shows the $H_T$ distributions for the
$e\tau_h$ and $\mu\tau_h$ channels, and includes the control region CR2J
and the two signal regions.

\begin{table}
\caption{Numbers of events observed in data and estimates for the
total background, for the $e\tau_h$ and $\mu\tau_h$ channels, in the
control regions (CR0J, CR1J, CR2J) and signal
regions (${\rm SR}_{\rm B}$, ${\rm SR}_{\rm A}$).
For the backgrounds, the statistical uncertainty is given first,
followed by the systematic uncertainty.}
\label{table:yields}
\begin{ruledtabular}
\begin{tabular*}{\hsize}{lc
@{\extracolsep{0ptplus1fil}}  D{,}{.}{-1}
c
@{\extracolsep{0ptplus1fil}}  D{,}{.}{-1}}
     & \multicolumn{2}{l}{\qquad \quad $e\tau_h$} & \multicolumn{2}{l}{\qquad \quad $\mu\tau_h$} \\
     & Data & \multicolumn{1}{c}{Background} & Data & \multicolumn{1}{c}{Background} \\ \colrule
\\[-2.5mm]
CR0J & $129$ & 122.1 \pm 2,1 \pm 11.3 & $129$ & 147.1 \pm 2,6 \pm 12.3 \\
CR1J & $110$ & 109.2 \pm 2,3 \pm 9.3 & $\phantom{0}79$ & 100.5 \pm 2,5 \pm 6.7 \\
CR2J & $\phantom{0}36$ & 33.4 \pm 1,4 \pm 4.8 & $\phantom{0}26$ & 30.6 \pm 1,6 \pm 3.8 \\
${\rm SR}_{\rm B}$ & $ \phantom{00}5 $ & 3.3 \pm 0,3 \pm 0.5 & $ \phantom{00}3 $ & 2.2 \pm 0,3 \pm 0.3 \\
${\rm SR}_{\rm A}$ & $ \phantom{00}0 $ & 0.3 \pm 0,1 \pm 0.1 & $ \phantom{00}0 $ & 0.2 \pm 0,1 \pm 0.1 \\
\end{tabular*}
\end{ruledtabular}
\end{table}

% Systematics

The dominant sources of systematic uncertainties on the signal efficiencies
are the amount of initial state radiation (ISR) and final state
radiation (FSR), the tau identification, and the isolation requirements.  
The ISR and FSR uncertainties, as evaluated by varying the amount of
ISR and FSR in simulation, are each approximately $3.7\%$ of the selection
efficiency.  The tau identification systematic uncertainty, as measured
using methods described elsewhere~\cite{ztautauprd}, is $3.0\%$ of the
selection efficiency and is
primarily due to uncertainties in the response of the hadronic calorimeter
to charged tracks and uncertainties in the track reconstruction efficiency.
Uncertainty in simulating the effect of the lepton isolation requirement
is evaluated by comparing the efficiency of this isolation requirement in data
and simulation.  This yields a $3.0\%$ uncertainty on the selection efficiency.
Additional sources of systematic uncertainty on the signal efficiency are
the choices of the PDFs and the jet energy scale,
as well as smaller contributions from the acceptance criteria, the
electron (muon) identification, and the $\met$ simulation.  The total
systematic uncertainty on signal efficiency ranges from $10.5\%$ ($10.4\%$)
for the $e\tau_h$ ($\mu\tau_h$) channel, for a \VLQ\ with mass
$\mVLQ = 160 \gevcc$, down to $7.0\%$ ($7.4\%$) for $\mVLQ = 400 \gevcc$.
The systematic uncertainties on the theoretical prediction of the \VLQ\ pair
production cross section arise from the choices of PDFs and
renormalization scale $Q^2$.  These two contributions are
combined in quadrature.

% Results

A likelihood function is constructed using a Poisson probability
distribution of the expected rates of signal plus backgrounds and observed
number of events in each channel and signal region.  For each \VLQ\ mass,
the expected signal rates are functions of the \VLQ\ pair production cross
section.  The expected signal rates also include factors for the branching
ratios of leptonic and hadronic tau decays, luminosity, and full selection
efficiencies.  Systematic uncertainties, including $6\%$ due to the luminosity
measurement (not included in Table~\ref{table:yields}), are incorporated
into the fit.  We apply Gaussian probabilities for the uncertainties on the
background estimates, and account for correlations among different sources
of systematic uncertainties.  To set cross section limits for each mass,
we integrate the likelihood distribution over all parameters except
the cross section, and then integrate as a function of cross section
from zero up to the cross section where the integral reaches $0.95$.

\begin{figure}[!ht]
\begin{center}
\includegraphics[width=8.5cm,angle=0]{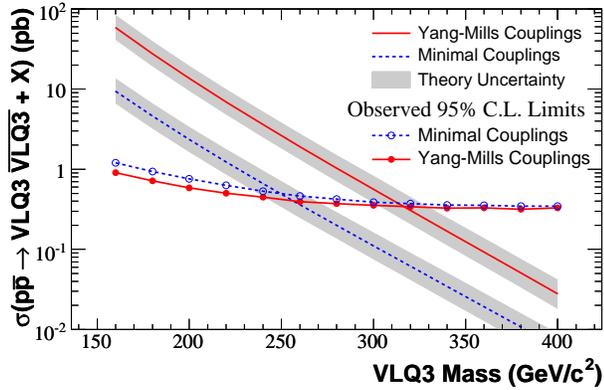}
\caption{The $95\%$ C.L.\ upper limits for \VLQ\ pair production versus
mass.  Also shown are the theoretical predictions using the simulation
described in the text, with bands for uncertainties due to the choices
of PDFs and $Q^2$.}
\label{fig:xsec_lim}
\end{center}
\end{figure}

The results are shown in Fig.~\ref{fig:xsec_lim}, as a function
of \VLQ\ mass, along with the theoretical predictions.  For a \VLQ\ with
Yang-Mills couplings, at $95\%$ confidence level (C.L.), the upper limit on
the cross section is $\sigma < 344 \fb$, assuming
${\cal B}(\VLQ \to\b\tau)=1$, and the lower limit on the mass is
$\mVLQ > 317 \gevcc$.  With theoretical uncertainties included on the
predicted cross section, the results are $\sigma < 360 \fb$ and
$\mVLQ > 294 \gevcc$.
For the minimal couplings model, the upper limit on the cross section is
$\sigma < 493 \fb$ and the lower limit on the mass is $\mVLQ > 251 \gevcc$.
With theoretical uncertainties included on the predicted cross section, the
results are $\sigma < 610 \fb$ and $\mVLQ > 223 \gevcc$.
The mass limits are approximately $80-90 \gevcc$ higher than those of
previous comparable results~\cite{CDFrunI,D0runI}.

% Conclusion

Using $322 \invpb$ of luminosity at CDF II, we have searched for \VLQ\ pair
production and subsequent decay to two tau leptons and two jets.  We observe
no excess of events beyond the expected SM processes and set the most
stringent limits to date on the \VLQ\ mass and pair production cross section
in the context of two coupling scenarios.

% from: http://www-cdf.fnal.gov/internal/physics/godparents/standardack.html
We thank the Fermilab staff and the technical staffs of the participating
institutions for their vital contributions.  This work was supported by the
U.S. Department of Energy and National Science Foundation; the Italian
Istituto Nazionale di Fisica Nucleare; the Ministry of Education, Culture,
Sports, Science and Technology of Japan; the Natural Sciences and Engineering
Research Council of Canada; the National Science Council of the Republic of
China; the Swiss National Science Foundation; the A.P. Sloan Foundation; the
Bundesministerium f\"ur Bildung und Forschung, Germany; the Korean Science
and Engineering Foundation and the Korean Research Foundation; the Science
and Technology Facilities Council and the Royal Society, UK; the Institut
National de Physique Nucleaire et Physique des Particules/CNRS; the Russian
Foundation for Basic Research; the Comisi\'on Interministerial de Ciencia y
Tecnolog\'{\i}a, Spain; the European Community's Human Potential Programme;
the Slovak R\&D Agency; and the Academy of Finland.


\begin{thebibliography}{99}
\bibitem{coords} We use a cylindrical coordinate system with the $z$ axis
along the beam axis, polar angle $\theta$, azimuthal angle $\phi$, and
pseudorapidity $\eta \equiv - \ln \tan(\theta/2)$.
We define $\pt \equiv p \sin\theta$ and $\Et \equiv E \sin\theta$, where
$p$ is the momentum measured by the tracking chambers and $E$ is the energy
measured by the calorimeters.
The missing transverse energy is defined by
$\vec{E}\!\!\!\!/_T = - \sum_i E_T^i \hat{n}_i$, where $i =$ calorimeter tower
number with $|\eta| < 3.6$ and $\hat{n}_i$ is a unit vector perpendicular
to the beam axis and pointing at the $i^{th}$ calorimeter tower.  We also
define $\met = |\vec{E}\!\!\!\!/_T|$.
\end{thebibliography}
\end{document}